\documentclass[twocolumn,aps,letter]{revtex4}
\usepackage{graphicx,graphics,float}
\usepackage{dcolumn}
\usepackage{bm}
\usepackage{amsmath,amssymb}
\usepackage{verbatim}
\usepackage{mathdots}
\usepackage{slashbox}
\begin{document}
\title{End-to-end correlations in the Kitaev chain}
\author{Jose Reslen}
\affiliation{Coordinaci\'on de F\'{\i}sica, Universidad del Atl\'antico, Kil\'ometro 7 Antigua v\'{\i}a a Puerto Colombia, A.A. 1890, Barranquilla, Colombia.}%
\date{\today}

\begin{abstract}
The interdependence between long range correlations and topological signatures
in fermionic arrays is examined. End-to-end correlations, in particular classical
correlations, maintain a characteristic pattern in the presence of delocalized
excitations and this behavior can be used as an operational criterion to identify
Majorana fermions in one-dimensional systems. The study discusses how to obtain 
the chain eigenstates in tensor-state representation together with the proposed 
assessment of correlations. Outstandingly, the final result can be written as a 
simple analytical expression that underlines the link with the system's 
topological phases.
%
\end{abstract}
\maketitle
\section{Introduction}
Majorana fermions \cite{eliot,alicea} are described by a highly versatile formalism that 
provides conceptual- as well as technical-tools, so much so that Majorana particles can be found 
in solid state systems incarnated as collective phenomena. One particular
scenario where this identification takes place is the Kitaev chain \cite{kitaev},
a one-dimensional fermionic system that displays a fundamental relation
between Majorana excitations and topology, allowing the use of geometrical
arguments to establish a connection
between the presence of Majorana fermions in the open chain and the
parity of the periodic chain, providing in this way simple and
operational conditions to allocate Majorana particles, specifically, superconductivity, 
which ensures a gap in the bulk, and an odd number of Dirac points in the band 
structure of the non-interacting system. This approach has been rather successful
in giving a qualitative characterization of the Majorana chain, and the
discovery has been attracting a lot of attention over the past decades
due to its potential applications in quantum information, 
prompting experimental verification in state-of-the-art setups, usually
in the form of zero-bias conductance peaks on the edges of one-dimensional 
structures, as for instance in \cite{deng,pawlak,sun} to mention only
the most recent studies.

Even though the Kitaev chain is well understood in terms of its 
topological structure, a more quantitative description in terms 
of the state's mean values is clearly of interest. Such a 
description is necessary to thoroughly characterize the behavior 
of the state's observables in the topological phase. This 
characterization can then be used on other systems where lack 
of integrability does not allow a direct identification of Majorana 
excitations \cite{gergs}. The fundamental
observation is that since the excitations supported by 
Majorana fermions are highly delocalized, it is reasonable to 
expect they enhance the correlations between the end sites of 
the chain. If that is indeed the case, experimental 
verification could be improved if it were possible to 
simultaneously test electron density on both of the chain 
ends. The study of edge correlations in fermion chains has been 
addressed in references \cite{miao1,wang} for specific
cases that permit analytical progress \cite{miao2,katsura}. 
Here the analysis covers the whole range of parameters and
is conceptually exact, albeit with a numerical component.    
This approach allows to find a generalized expression for
the correlations that complements the results derived
analytically.

The fact that the Kitaev chain is an integrable model does 
not preclude the need for numerical analysis. Complications 
arise because the chain's Hilbert space grows as $2^N$, 
being $N$ the number of sites, and many features manifest 
exclusively in the thermodynamic limit. These complications 
can be circumvented by the use of tensor state techniques. 
Such techniques can be implemented in different ways. One 
such way is Density Matrix Renormalization Group 
(DMRG) \cite{miao1,gergs,kitaev2}, which minimizes 
the energy over the space of eigenstates of the chain's local 
density matrices. Another approach is to use a tensor
representation as a variational network in an abstract way. 
This is characteristic of
the method known as Matrix Product States (MPS). The way tensor
state techniques are applied here is different from these two, and
is more in accordance with the updating protocols introduced in
reference \cite{vidal}. A family of methods based on such
protocols is known as Time Evolving Block Decimation (TEBD). However,
the path followed in this report differs from this denomination. First, 
time-evolving- or step-integration is not incorporated, and second, the implementation
is exact, so that numerical approximations like splittings of operator exponentials, which
are huge error-contributors to TEBD, are not employed whatsoever. The techniques
applied here to fermion chains have been first developed
in the context of bosonic arrays in \cite{ReslenRMF,ReslenIOP},
although with some key differences, the most important one being the
inclusion of pairing terms integrally in the current formalism, which is possible thanks
to the decomposition of fermionic operators in terms of Majorana
operators. Another aspect that contrasts with other works is that the tensor state formulation
is carried completely in the fermionic Fock-space, without the extra work
of incorporating the so called Wigner-Jordan transformation to reformulate the
problem in terms of a spin chain, as seems to be frequent in DMRG 
applications to fermion systems.

The Kitaev chain, also known as the Majorana chain, is described by the following Hamiltonian \cite{kitaev}
\begin{gather}
\hat{H} = \sum_{j=1}^N -w ( \hat{c}^{\dagger}_j \hat{c}_{j+1} + \hat{c}^{\dagger}_{j+1} \hat{c}_j ) - \mu \left( \hat{c}^{\dagger}_j \hat{c}_{j} - \frac{1}{2} \right ) \nonumber \\
+ \Delta \hat{c}_j \hat{c}_{j+1} + \Delta^* \hat{c}^{\dagger}_{j+1} \hat{c}^{\dagger}_j.
\label{kita}
\end{gather}
Constant $w$ is the next-neighbor hopping intensity, while $\mu$ is the 
chemical potential, which relates to the total number of fermions in the wire.
Parameter $\Delta$ is the intensity of the pairing and is known as the 
superconducting gap.
Creation and annihilation operators follow fermionic
anticommutation rules $\{\hat{c}_j,\hat{c}_k\} = 0$ and 
$\{\hat{c}_j,\hat{c}_k^\dagger\} = \delta_j^k$. Open or periodic boundary 
conditions are enforced by taking $\hat{c}_{N+1}=0$ or $\hat{c}_{N+1}=\hat{c}_1$
respectively. The model describes a quantum wire in proximity to the surface of 
a p-wave superconductor \cite{kitaev,eliot,alicea}, and also a spin-polarized 1-D
superconductor, since only one spin component is being considered and 
the pairing term mixes modes with opposite crystal momentum,
as can be shown by switching to a momentum basis. Independently of boundary conditions, 
Hamiltonian (\ref{kita}) commutes with the parity operator
\begin{gather}
\hat{\Pi} = e^{i \pi \sum_{j=1}^{N} \hat{c}_j^{\dagger} \hat{c}_j}.
\label{rainy}
\end{gather}
The symmetry associated to this operator is not spatial, instead, it
is a parity associated to number of particles. Let us now introduce 
the Majorana operators (MOs) corresponding to site $j$:
\begin{gather}
\hat{\gamma}_{2 j - 1} = e^{\frac{i\phi}{2 }} \hat{c}_j + e^{-\frac{i\phi}{2 }} \hat{c}^{\dagger}_j, \label{eight} \\
\hat{\gamma}_{2 j} = - i e^{\frac{i\phi}{2 }} \hat{c}_j + i e^{-\frac{i\phi}{2 }} \hat{c}^{\dagger}_j. \label{nine}
\end{gather}
A key feature of these MOs is that they are hermitian,
$\hat{\gamma}_k^\dagger = \hat{\gamma}_k$. It can be shown that the anticommutation relations
are given by $\{ \hat{\gamma}_k, \hat{\gamma}_j \} = 2 \delta_k^j$, so that 
$\hat{\gamma}_k^2 = 1$. Since there are two Majorana modes for every site, the
total number of modes doubles. Equations (\ref{eight}) and (\ref{nine})
can be inverted and the result can be used to write Hamiltonian (\ref{kita})
as
\begin{gather}
\hat{H} = \frac{i}{2}\sum_{j=1}^N \left ( -\mu \hat{\gamma}_{2j-1} \hat{\gamma}_{2j} + (|\Delta|-w) \hat{\gamma}_{2j-1} \hat{\gamma}_{2j+2} \right . \nonumber \\
\left . + (|\Delta|+w) \hat{\gamma}_{2j} \hat{\gamma}_{2j+1} \right ).
\label{kike}
\end{gather}
When $w = \Delta = 0$, this Hamiltonian becomes diagonal in the Fock basis with
a non-degenerate spectrum and a ground state that depends on the sign of $\mu$. 
From (\ref{kike}) it can be seen that such a particular case corresponds to a chain 
where MOs from the same site pair up. This behavior is the generic signature of the 
{\it trivial phase}. In contrast, when 
$w = |\Delta| > 0$ and $\mu =0$, the pairing takes place between Majorana
operators from neighbor sites, as can be seen in the transformed Hamiltonian  
\begin{gather}
\hat{H} = i w \sum_{j=1}^{N-1} \hat{\gamma}_{2j} \hat{\gamma}_{2j+1}.
\label{aqua}
\end{gather}
A key feature of this expression is that it lacks both $\hat{\gamma}_{1}$ and 
$\hat{\gamma}_{2N}$, which are {\it unpaired}. From these one can build an uncoupled mode, 
\begin{gather}
\hat{f}_N = \frac{1}{2} \left( \hat{\gamma}_{1} + i \hat{\gamma}_{2 N} \right),
\end{gather}
satisfying $\{\hat{f}_N, \hat{f}_N^\dagger\} = 1$. This is 
a highly delocalized fermionic mode, having equitable contributions on the edges. 
The simplest physical operator that can be in this way built 
is $\hat{f}_N^\dagger \hat{f}_N$, which consequently commutes with the 
Hamiltonian. The Hilbert space associated to such a term contains two 
states, one occupied and one empty. As the Hamiltonian does not include 
terms that could operate on neither of these, it is energetically equivalent
to have many-body eigenstates with or without a particle on the 
aforementioned mode, i.e., the energy cost associated to this mode
is zero, being that the reason why $\hat{f}_N$ is known as a Zero Mode
while $\hat{\gamma}_{1}$ and $\hat{\gamma}_{2N}$ are known as 
Majorana Zero Modes (MZMs) \cite{eliot} or Edge Modes. As a consequence, the whole 
spectrum of (\ref{aqua}) becomes two-fold degenerate. The fact that 
neither $\hat{\gamma}_{1}$ nor $\hat{\gamma}_{2N}$ appear in the 
Hamiltonian implies that they do not pick up oscillatory phases
in the Heisenberg picture, which makes them robust against this kind
of decoherence mechanism. From the previous arguments it can be deduced 
that (\ref{aqua}) commutes with the symmetry operator
\begin{gather}
\hat{Q}_R = i \hat{\gamma}_1 \hat{\gamma}_{2 N} = 2 \hat{f}_N^\dagger \hat{f}_N - 1.
\label{delmar}
\end{gather}
It can be noticed that $\hat{Q}_R$ is unitary and its eigenvalues are
$1$ for a filled mode and $-1$ for an unfilled one. Unlike $\hat{\Pi}$,
$\hat{Q}_R$ determines a symmetry only for a specific set of parameters.
In spite of the spectrum being degenerate in this case, it is possible 
to build ground states $|\psi_G\rangle$ of (\ref{aqua}) that are also 
eigenstates of (\ref{delmar}), so that
\begin{gather}
|\langle \psi_G,  \hat{Q}_R \psi_G \rangle| = 1.
\label{crowded_house}
\end{gather}
Performing the same calculation with any normalized state that is not
an eigenstate of $\hat{Q}_R$ would result in a lower value, so that 
the maximum is linked to eigenstates of Hamiltonians having unpaired 
MOs completely localized at the ends. A totally analogous
case would be obtained if focus is made on the point 
$w = -|\Delta| < 0$ and $\mu =0$, yielding
\begin{gather}
\hat{H} = - i w \sum_{j=1}^{N-1}  \hat{\gamma}_{2j-1} \hat{\gamma}_{2j+2}.
\end{gather}
This time it is $\hat{\gamma}_{2}$ and $\hat{\gamma}_{2N-1}$ which do not
appear in the Hamiltonian, thus giving rise to the symmetry operator
\begin{gather}
\hat{Q}_L = i \hat{\gamma}_2 \hat{\gamma}_{2 N-1},
\end{gather}
which follows a relation analogous to (\ref{crowded_house}). According 
to reference \cite{kitaev}, there are unpaired MOs, or Majorana fermions, over the region 
in parameter space surrounding the particular cases studied above as 
long as the gap of the equivalent periodic chain does not vanish. Following
this argument it can be shown that unpaired MOs prevail in the region
defined by $2 |w| < |\mu|$. In the general case such operators only 
become completely unpaired in the thermodynamic limit, although with exponential convergence, 
and they are not completely, but still highly, localized at the ends. 

The above observations suggest that in order to scan for 
unpaired MOs it is useful to exploit their relation with the 
state's local-symmetries. Let us therefore define the operator
\begin{gather}
\hat{Q} = \hat{Q}_L + \hat{Q}_R = 2 (\hat{c}_1 \hat{c}_N^\dagger + \hat{c}_N \hat{c}_1^\dagger). 
\end{gather}
It is reasonable to expect that the mean values of $\hat{Q}$, which
actually measure end-to-end single-particle hopping, determine the degree of localization
of MOs at the chain ends, taking extreme values when
they are completely localized and vanishing when there is none. In order
to test this conjecture the following measure is proposed
\begin{gather}
Z = \lim_{N\rightarrow \infty} |\langle \hat{Q} \rangle|.    
\label{terror}
\end{gather}
The numerical calculation of this expression is not always efficient because 
long-range correlations are involved, hence a practical approach is desirable. 
Next section focuses on presenting a way in which the eigenstates of (\ref{kita}) 
as well as $Z$ can be effectively calculated.
\section{Reduction of the Kitaev chain by a series of unitary transformations}
\label{aji}
Hamiltonian (\ref{kike}) has the following general structure
\begin{gather}
\hat{H} = \frac{i}{4} \sum_{k=1}^{2N} \sum_{l=1}^{2N} A_{kl} \hat{\gamma}_k \hat{\gamma}_l,
\label{twelve}
\end{gather}
where the coefficients $A_{kl}$ form a real antisymmetric matrix $A_{kl} = -A_{lk}$. 
Following Kitaev \cite{kitaev}, $\hat{H}$ can be diagonalized by 
an unitary transformation that reduces it to
{\scriptsize
\begin{gather}
\hat{H} = \frac{i}{4} \sum_{k=1}^{N} \epsilon_k \left( \hat{\zeta}_{2k-1} \hat{\zeta}_{2k} -  \hat{\zeta}_{2k} \hat{\zeta}_{2k-1} \right) = \frac{i}{2} \sum_{k=1}^{N} \epsilon_k \hat{\zeta}_{2k-1} \hat{\zeta}_{2k}.
\label{thirdteen}
\end{gather}
}
The $\hat{\zeta}$'s are MO that can be expressed
as linear combinations of the $\hat{\gamma}$'s 
\begin{gather}
\left (
\begin{array}{c}
\zeta_1  \\
\zeta_2  \\
\zeta_3  \\
\zeta_4 \\
\vdots
\end{array}
\right ) =
\hat{W}
\left (
\begin{array}{c}
\gamma_1  \\
\gamma_2  \\
\gamma_3  \\
\gamma_4 \\
\vdots
\end{array}
\right ).
\end{gather}
Matrix $\hat{W}$ is such that it transform $\hat{A}$ in the following manner
\begin{gather}
\hat{W} \hat{A} \hat{W}^\dagger =
\left (
\begin{array}{ccccc}
0 & \epsilon_1 & 0 & 0 & \hdots  \\
-\epsilon_1 & 0 & 0 & 0 & \hdots  \\
0 & 0 &  0 & \epsilon_2 & \hdots  \\
0 & 0 & -\epsilon_2  & 0 & \hdots  \\
\vdots & \vdots & \vdots  & \vdots & \ddots   
\end{array}
\right ). 
\label{sixteen}
\end{gather}
The single-body energies $\epsilon_k$ are real-and-positive while $\hat{W}$
is a real orthonormal matrix satisfying $\hat{W} \hat{W}^\dagger = I$.
This factorization is a particular case of a procedure known as Schur decomposition \cite{SD}.
The diagonalized Hamiltonian can be written in terms of standard fermionic
modes 
\begin{gather}
\hat{H} = \sum_{k=1}^N \epsilon_k \left( \hat{f}_k^\dagger \hat{f}_k - \frac{1}{2} \right), \text{ } \hat{f}_k = \frac{1}{2}(\hat{\zeta}_{2k-1} + i \hat{\zeta}_{2k}).
\label{fourteen}
\end{gather}
This can be checked by replacing $\hat{\zeta}_{2 k - 1} =  \hat{f}_k +  \hat{f}^{\dagger}_k$
and $\hat{\zeta}_{2 k} = - i  \hat{f}_k + i \hat{f}^{\dagger}_k$ in (\ref{thirdteen}).  
The system eigenenergies are given by
\begin{gather}
E_l = \sum_{k=1}^{N} \epsilon_k \left(n_k - \frac{1}{2} \right),
\label{pinky}
\end{gather}
in such a way that $n_k = \{0,1\}$. The system's ground state corresponds to the case where all $n_k=0$,
therefore $E_G = -\sum_{k=1}^{N} \frac{\epsilon_k}{2}$. If there are one or 
more single-body vanishing-energies $\epsilon_k = 0$, the spectrum becomes degenerate, 
because there is no energy difference between occupied and unoccupied zero modes.

In order to obtain the eigenstates an approach similar to that
in reference \cite{wagner} is adopted. First, Hamiltonian (\ref{fourteen}) is written as 
{\scriptsize
\begin{gather}
\hat{H} = \sum_{k=1}^N \epsilon_k \left( \frac{1}{4} \left( \sum_{j=1}^{2N} W_{2k-1,j}\hat{\gamma}_{j} - i \sum_{j=1}^{2N} W_{2k,j} \hat{\gamma}_{j} \right) \right . \times \nonumber \\ 
\left . \left( \sum_{j=1}^{2N} W_{2k-1,j}\hat{\gamma}_{j} + i \sum_{j=1}^{2N} W_{2k,j} \hat{\gamma}_{j} \right)     - \frac{1}{2} \right).
\end{gather}
}
The $W_{j,k}$'s are the components of matrix $\hat{W}$, as defined by (\ref{sixteen}).
An unitary transformation acting on two consecutive MOs can be defined as
\begin{gather}
\left . \hat{U}^{[j]} \right .^{-1} = e^{\frac{\theta}{2} \hat{\gamma_j} \hat{\gamma}_{j-1}}.
\end{gather}
The effect of this transformation on the Hamiltonian is calculated through 
$\hat {H} \rightarrow \left . \hat{U}^{[j]} \right .^{-1} \hat{H}  \hat{U}^{[j]}$.
This can be performed by applying the same transformation on every
operator composing $\hat{H}$. The operation over a linear combination of
consecutive MOs is written as
{\small
\begin{gather}
\left . \hat{U}^{[j]} \right .^{-1} ( W_{j-1} \hat{\gamma}_{j-1} + W_{j} \hat{\gamma}_{j} ) \hat{U}^{[j]} = W_{j}' \hat{\gamma}_{j} + W_{j-1}' \hat{\gamma}_{j-1},
\end{gather}
}
where $W_{j}'= W_{j} \cos \theta - W_{j-1} \sin \theta$ and $W_{j-1}' = W_{j-1} \cos \theta + W_j \sin \theta$.
The contribution of $\hat{\gamma}_j$ can be taken out by choosing
\begin{gather}
\tan \theta = \frac{W_j}{W_{j-1}},
\label{fragance}
\end{gather}
in such a way that $W_j'=0$. If $W_{j-1} = 0$ and $W_{j} \ne 0$, then $\theta=\frac{\pi}{2}$. If both
$W_{j-1}$ and $W_{j}$ are zero, then $\theta=0$ is enforced. In any other case the angle
is well defined because the $W$'s are real. It is practical to choose the 
angle $\theta$ so that $sign(\sin \theta) = sign(W_j)$ and $sign(\cos \theta) = sign(W_{j-1})$.
In this way $W_{j-1}' = \sqrt{W_{j-1}^2 + W_{j}^2}$, leaving a positive coefficient. 
In order to highlight the dependence of $\theta$ with respect to $W_j$ and $W_{j-1}$, 
$\theta_j$ is used from now on. When this operation is applied on the whole Hamiltonian, 
the mechanism can be described as an overall action on the diagonal modes:
{\tiny
\begin{gather}
\begin{array}{c|c|c}  
& \left . \hat{U}_1^{[2N]} \right .^{-1}  &  \\ \cline{2-2}
\hat{\zeta}_1 = & W_{1,2N}\hat{\gamma}_{2N} + W_{1,2N-1} \hat{\gamma}_{2N-1} & + W_{1,2N-2} \hat{\gamma}_{2N-2}  \dots + W_{1,1} \hat{\gamma}_{1} \\ 
\hat{\zeta}_2 = & W_{2,2N}\hat{\gamma}_{2N} + W_{2,2N-1} \hat{\gamma}_{2N-1} & + W_{2,2N-2} \hat{\gamma}_{2N-2}  \dots + W_{2,1} \hat{\gamma}_{1} \\ 
\vdots & \vdots  &   \vdots \\
\end{array} \nonumber
\end{gather}
}
As a result, $\hat{\gamma}_{2N}$ vanishes from $\hat{\zeta}_1$ and the vertically 
aligned coefficients are in some way affected. The process continues
by applying another transformation aimed at canceling the next component, which generates
a similar effect on the stack of coefficients
{\scriptsize
\begin{gather}
\begin{array}{c|c|c}  
& \left . \hat{U}_1^{[2N-1]} \right .^{-1} &  \\ \cline{2-2}
 & W_{1,2N-1}'\hat{\gamma}_{2N-1} + W_{1,2N-2} \hat{\gamma}_{2N-2} &  \dots + W_{1,1} \hat{\gamma}_1 \\ 
 W_{2,2N}'\hat{\gamma}_{2N} + & W_{2,2N-1}'\hat{\gamma}_{2N-1} + W_{2,2N-2} \hat{\gamma}_{2N-2} &  \dots + W_{2,1} \hat{\gamma}_1 \\
\vdots & \vdots & \vdots \\
W_{2N,2N}'\hat{\gamma}_{2N} + & W_{2N,2N-1}'\hat{\gamma}_{2N-1} + W_{2N,2N-2} \hat{\gamma}_{2N-2} &  \dots + W_{2N,1} \hat{\gamma}_1  \\  
\end{array} \nonumber 
\end{gather}
}
The process is repeated, removing one component in each step and advancing toward $\hat{\gamma}_1$
{\scriptsize
\begin{gather}
\begin{array}{c|c|}  
& \left . \hat{U}_1^{[2]} \right .^{-1}  \\ \cline{2-2}
   & W_{1,2}'\hat{\gamma}_{2} + W_{1,1} \hat{\gamma}_{1}  \\ 
W_{2,2N}'\hat{\gamma}_{2N} + \dots + W_{2,3}''\hat{\gamma}_{3} + & W_{2,2}'\hat{\gamma}_{2} + W_{2,1} \hat{\gamma}_{1}  \\
\vdots & \vdots  \\
 W_{2N,2N}'\hat{\gamma}_{2N} + \dots + W_{2N,3}''\hat{\gamma}_{3} + & W_{2N,2}'\hat{\gamma}_{2} + W_{2N,1} \hat{\gamma}_{1}  \\  
\end{array} \nonumber
\end{gather}
}
The last transformation eliminates $\hat{\gamma}_2$ and leaves only $\hat{\gamma}_1$
multiplied by $ W_{1,1}' = \sqrt{W_{1,1}^2 + W_{1,2}^2 + ... + W_{1,2N}^2} = 1$.
Because all the transformations are unitary, the orthogonality of the coefficients must be preserved. 
Hence, if only $\hat{\gamma}_1$ remains in the top row, there cannot be $\hat{\gamma}_1$-terms
in the rest of the stack. The last operation then leaves the following arrangement
\begin{gather}
\begin{array}{cc}  
   & \hat{\gamma}_{1}  \\ 
W_{2,2N}'\hat{\gamma}_{2N} + \dots + W_{2,3}''\hat{\gamma}_{3} + W_{2,2}''\hat{\gamma}_{2}  &   \\
\vdots &  \\
W_{2N,2N}'\hat{\gamma}_{2N} + \dots + W_{2N,3}''\hat{\gamma}_{3} + W_{2N,2}''\hat{\gamma}_{2}  &   \\  
\end{array} \nonumber
\end{gather}
A similar series of operations can be devised to reduce the second row, this time
avoiding any transformation involving $\hat{\gamma}_1$ in order to keep
the first mode folded. The process can be repeated with the
same intended effect in each step. However, the last transformation brings up an additional issue. 
Let us notice that before the last operation the stack of components looks like
\begin{gather}
\begin{array}{|c|cccc}  
\left . \hat{U}_{2N-1}^{[N]} \right .^{-1} & &  & &  \\ \cline{1-1}
 & & & & \hat{\gamma}_{1}  \\ 
 & & & \hat{\gamma}_{2} &  \\ 
 & \iddots &  & &  \\ 
 W_{2N-1,2N}\hat{\gamma}_{2N} + W_{2N-1,2N-1}\hat{\gamma}_{2N-1}   & & & &  \\
W_{2N,2N}\hat{\gamma}_{2N} + W_{2N,2N-1}\hat{\gamma}_{2N-1}  & &  & &   \\  
\end{array} \nonumber
\end{gather}
Transformation $\left . \hat{U}_{2N-1}^{[N]} \right .^{-1}$ is aimed at folding the antepenultimate row, however, due to
the orthonormality of the original modes, it folds the last row too. However, 
there is no guarantee that the resulting coefficient is positive, since the transformation
only takes care of the sign of the coefficients of the row being folded. Consequently, after
the folding is finished, there are two possible states of the stack
\begin{gather}
\begin{array}{|ccc|}  
& & \hat{\gamma}_{1}  \\ 
& \iddots  &  \\ 
\hat{\gamma}_{2N} & &    \\  
\end{array} 
\hspace{1cm}
\text{ or }
\hspace{1cm}
\begin{array}{|ccc|}  
& & \hat{\gamma}_{1}  \\ 
& \iddots  &  \\ 
-\hat{\gamma}_{2N} & &    \\  
\end{array} \nonumber 
\end{gather}
In the first case, when all the coefficients are positive, the transformed Hamiltonian becomes
\begin{gather}
\sum_{k=1}^N \epsilon_k \left( \frac{1}{4} \left( \gamma_{2k-1} - i \gamma_{2k} \right)    \left( \gamma_{2k-1} + i \gamma_{2k} \right) - \frac{1}{2} \right) \nonumber \\ 
= \sum_{k=1}^N \epsilon_k \left( \hat{c}_k^\dagger \hat{c}_k - \frac{1}{2} \right).
\end{gather}
The eigenstates of this Hamiltonian can be identified as occupation states, 
\begin{gather}
|\varphi_l \rangle = \prod_{k=1}^{N} \left( \hat{c}_k^\dagger \right )^{n_k} |0\rangle,
\label{seventeen}
\end{gather}
and the corresponding eigenergies are given by (\ref{pinky}).
The vacuum $|0\rangle$, or state without fermions, is simultaneously
the system's ground state.

With regard to the second case, let us first point out that fermionic
operators are given in terms of MO by the relation
\begin{gather}
\hat{c}_{j} = \frac{e^{-\frac{i\phi}{2}}}{2} \left( \hat{\gamma}_{2j-1} + i \hat{\gamma}_{2 j} \right), \text{ }
\hat{c}_{j}^\dagger = \frac{e^{\frac{i\phi}{2}}}{2} \left( \hat{\gamma}_{2j-1} - i \hat{\gamma}_{2 j} \right). \nonumber
\end{gather}
It can be seen that a negative sign in $\hat{\gamma}_{2N}$ induces an particle-hole 
transformation, $\hat{c}_N^\dagger \Leftrightarrow \hat{c}_N$, in such a way that 
after passing to the fermionic basis the reduced Hamiltonian becomes
\begin{gather}
\sum_{k=1}^{N-1} \epsilon_k \left( \hat{c}_k^\dagger \hat{c}_k - \frac{1}{2} \right) - \epsilon_N \left( \hat{c}_N^\dagger \hat{c}_N - \frac{1}{2} \right).
\end{gather}
The eigenstates of this reduced Hamiltonian can be built as in (\ref{seventeen}), but
taking into account that the ground state is not the vacuum but the state with one fermion
in the $N$-th site
\begin{gather}
|\varphi_l \rangle = \prod_{k=1}^{N-1} \left( \hat{c}_k^\dagger \right )^{n_k} \hat{c}_N^{n_N}  |0...01\rangle.
\label{zeera}
\end{gather}
Likewise, the expression for the associated eigenenergy is (\ref{pinky}). 
To obtain the eigenstates of the original Kitaev chain, $| \psi_l \rangle$, one applies the 
transformations in reverse order over the states (\ref{seventeen}) or (\ref{zeera}), depending
on the result of the folding. The operation can be written as
\begin{gather}
| \psi_l \rangle = \prod_{k=2N-1}^1 \left ( \prod_{j=k+1}^{2N} \hat{U}_k^{[j]} \right ) | \varphi_l \rangle.
\label{twenty}
\end{gather}
Both $| \psi_l \rangle$ and $| \varphi_l \rangle$ are eigenstates corresponding
to $E_l$, because unitary transformations do not change eigenvalues. 
Using (\ref{eight}) and (\ref{nine}) it can be shown that the transformations 
that appear in (\ref{twenty}) are given by
\begin{gather}
\hat{U}_k^{[j]} = e^{-i\theta_{j,k} \left ( \hat{c}_{\frac{j}{2}}^\dagger \hat{c}_{\frac{j}{2}} - \frac{1}{2} \right )}, \text{if $j$ is even},
\end{gather}
and
\begin{gather}
\hat{U}_k^{[j]} = \exp \left [\frac{i \theta_{j,k}}{2}  \left ( -e^{\frac{i\phi}{2}}  \hat{c}_{\frac{j-1}{2}} + e^{\frac{-i\phi}{2}} \hat{c}_{\frac{j-1}{2}}^\dagger   \right ) \right . \times \nonumber \\
\left . \left ( e^{\frac{i\phi}{2}}  \hat{c}_{\frac{j+1}{2}} + e^{\frac{-i\phi}{2}} \hat{c}_{\frac{j+1}{2}}^\dagger   \right )  \right ], \text{if $j$ is odd}.
\end{gather}
Transformations with $j$ even operate only on site $\frac{j}{2}$ and in
matrix form they can be written as
\begin{gather}
\hat{U}_k^{[j]} = 
\left (
\begin{array}{cc}
 e^{\frac{i\theta_{j,k}}{2}}  &               0               \\
              0               & e^{-\frac{i\theta_{j,k}}{2}}  \\
\end{array}
\right ).
\label{twenty-one}
\end{gather}
This matrix is written with respect to occupation states with the
order $|0\rangle, |1\rangle$. Transformations with $j$ odd operate 
non-trivially on consecutive sites $\frac{j-1}{2}$ and $\frac{j+1}{2}$  
through the following matrix representation
\begin{gather}
\hat{U}_k^{[j]} = 
\left (
\begin{array}{cccc}
 \cos{\frac{\theta_{j,k}}{2}}  &               0                 &               0                   &   i\sin{\frac{\theta_{j,k}}{2}}  \\
              0                &  \cos{\frac{\theta_{j,k}}{2}}   &   i\sin{\frac{\theta_{j,k}}{2}}   &                0                 \\
              0                &  i\sin{\frac{\theta_{j,k}}{2}}  &   \cos{\frac{\theta_{j,k}}{2}}    &                0                 \\
i\sin{\frac{\theta_{j,k}}{2}}  &               0                 &               0                   &  \cos{\frac{\theta_{j,k}}{2}}   
\end{array}
\right ).
\label{twenty-two}
\end{gather}
In this case the basis order is $|00\rangle, |01\rangle, |10\rangle, |11\rangle$
(the first position for site $\frac{j-1}{2}$ and the second for site $\frac{j+1}{2}$).
The reducibility of the matrix underlines the fact that the Hamiltonian commutes
with the parity operator (\ref{rainy}) and therefore the eigenvectors inhabit spaces
with even or odd parity. Because these matrices only mix states
with the same parity, $| \psi_l \rangle$ and $| \varphi_l \rangle$ have equal
parity.

In order to obtain the eigenstates of the Kitaev chain, first matrix $\hat{W}$ is 
gotten using standard numerical routines. The entries of this matrix are then used 
to get the folding angles $\theta_{j,k}$ from Eq. (\ref{fragance}). These angles 
are then employed to build the transformations composing expression (\ref{twenty}) 
in matricial form. Since such transformations involve neighbor sites only, 
expression (\ref{twenty}) can be computed using the updating protocols described
in \cite{vidal}, so that the final result is expressed in tensorial
representation. A detailed description of how to incorporate tensor
product tecniques particularly on this problem is given in appendix \ref{tensor}.
\section{Results}
\label{tado}
\begin{figure}[h]
\begin{center}
\includegraphics[width=0.3\textwidth,angle=-90]{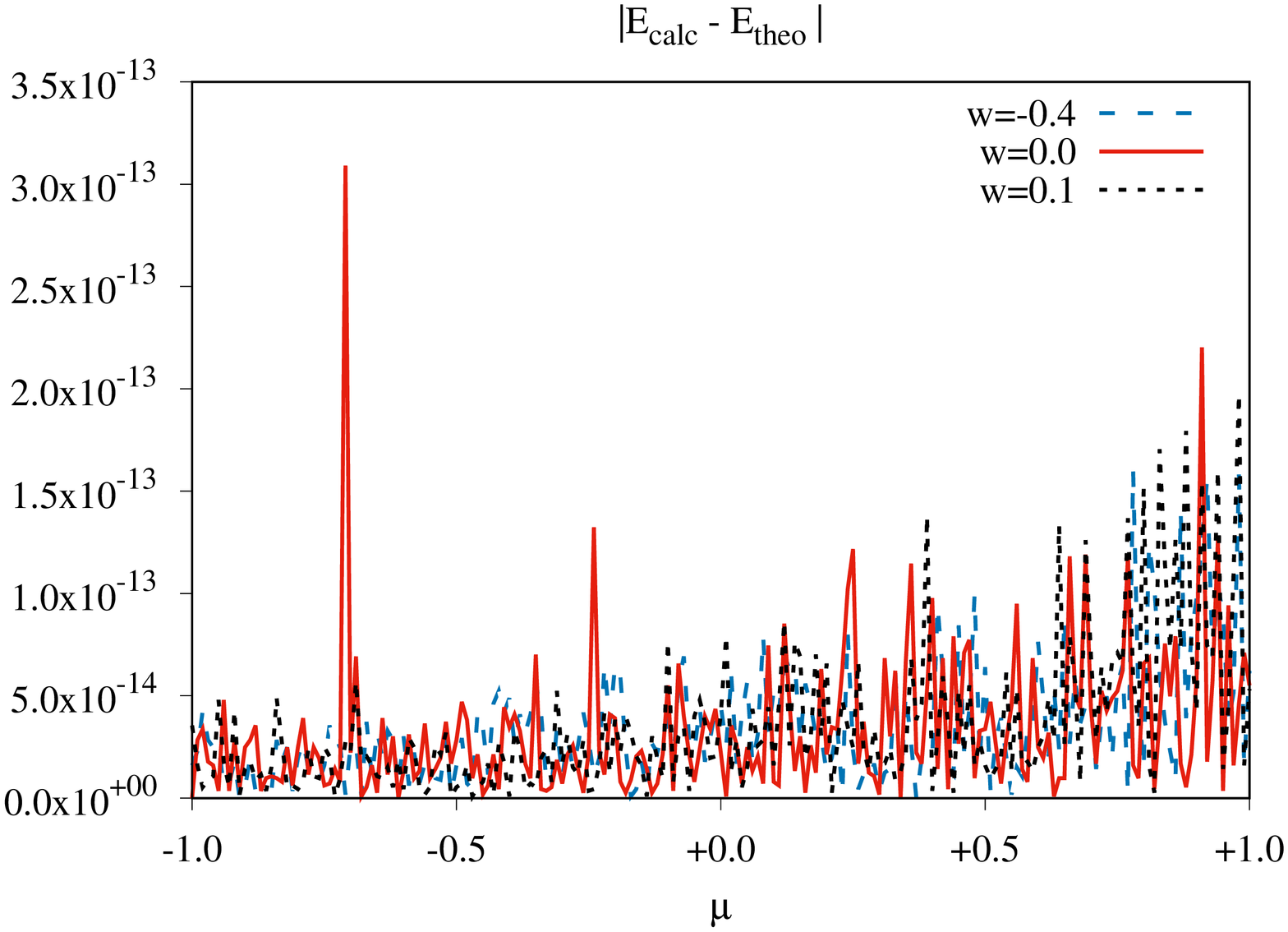} \\
\includegraphics[width=0.3\textwidth,angle=-90]{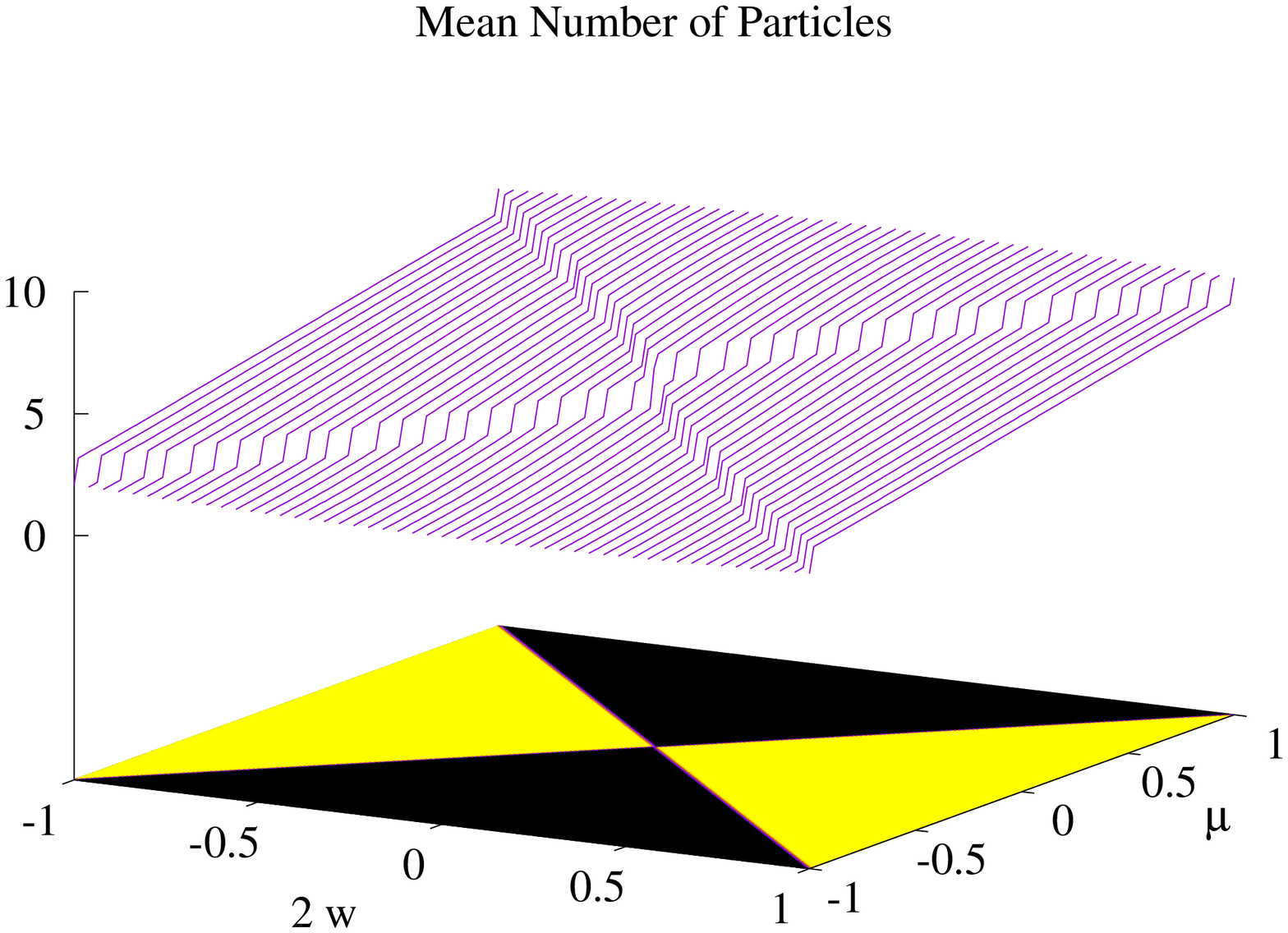}
\caption{Top. Accuracy of the ground-state energy in a periodic chain 
as delivered by the method described in section \ref{aji}. Bottom. Mean number 
of particles for the ground state of a periodic chain. The plot floor 
highlights the zones where the ground state parity is even (black) and
odd (yellow), the latter case also corresponds to the region where the
open chain holds Majorana fermions. In both figures $N=10$ and $\Delta=1$.}    
\label{fig6}
\end{center}
\end{figure}

Before addressing the study of correlations in the open chain, the
numerical approach proposed in the previous section is tested using
the spectrum of the periodic chain. The single-body energies are given by
{\footnotesize
\begin{gather}
E_{k}^{\pm} = \pm \sqrt{ \left( 2 w \cos \left( \frac{2\pi k}{N} \right) + \mu \right)^2 + 4 |\Delta|^2 \sin^2 \left( \frac{2\pi k}{N} \right)},
\end{gather}
}
for $1 \le k < \frac{N}{2}$. Additionally, $E_{\frac{N}{2}}^{+} = 2 w - \mu$
and $E_{\frac{N}{2}}^{-} = -2 w - \mu$. Such energies
are numerically calculated as a by-product of the Schur decomposition in 
(\ref{sixteen}) and then compared against these analytical results.   
Next, Eq. (\ref{pinky}) is used to find the ground state energy. 
The tensorial representation of the ground 
state is then obtained following the folding protocol discussed before. The energy 
of such a ground state is calculated {\it from this tensorial
representation}. This can be done as $N$ times the energy 
of two consecutive sites, but only for eigenstates with translational symmetry,
which is the case as long as such eigenstates are nondegenerate. 
This result is then compared with the quantity obtained before as the sum of 
single-body-energies. The top plot in figure \ref{fig6} shows the absolute
difference between the two estimates.   
The bottom plot depicts the mean number of particles, showing a 
behavior consistent with the contribution of a zero mode at $|\mu| = 2 |w|$.

Having verified the technique, let us now study correlations in 
the open chain. If the ground state is nondegenerate, $Z$ 
in (\ref{terror}) can be determined from
\begin{gather}
Z = \lim_{N\rightarrow \infty}| Tr(\hat{\rho}_{1N} \hat{Q})|,
\label{ecopetrol}
\end{gather}
where $\hat{\rho}_{1N}$ is the reduced density matrix of the chain 
ends. The computation of such a matrix from a state written in
tensorial representation is described in appendix \ref{rdm}. Matrix
$\hat{Q}$ comes given by
\begin{gather}
\hat{Q} =
\left (
\begin{array}{cccc}
0       & 0       & 0       & 0 \\
0       & 0       & 2 (-1)^P  & 0 \\
0       & 2 (-1)^P  & 0       & 0 \\
0       & 0       & 0       & 0
\end{array}
\right),
\end{gather}
where $P$ is the ground state's parity. $Z$ is found as the saturation 
value of (\ref{ecopetrol}) with respect to $N$. The results are shown in 
table \ref{cervanto}. The signs of $w$ and $\mu$ do not seem to 
influence the outcome. The observed behavior is compatible with
the notion that $Z$ is correlated to the contribution of unpaired MOs. The maxima 
are located at points with total localization of MOs and vanishing values
characterize the trivial phase. One would expect that
the non vanishing values of $Z$ provide an assessment of the level 
of localization of Majorana fermions at the edges.
\begin{table}[h]
{\scriptsize
\begin{center}
\begin{tabular}{|c|c|c|c|c|c|c|c|c|c|} \hline
\backslashbox{$\mu$}{$2 w$} &  -4   &  -3   &  -2   &  -1   &   0   &   1   &   2   &   3   &   4    \\ \hline
  4                         &   -   & 0.000 & 0.000 & 0.000 & 0.000 & 0.000 & 0.000 & 0.000 &   -    \\ \hline
  3                         & 0.388 &   -   & 0.000 & 0.000 & 0.000 & 0.000 & 0.000 &   -   & 0.388  \\ \hline
  2                         & 0.666 & 0.533 &   -   & 0.000 & 0.000 & 0.000 &   -   & 0.533 & 0.666  \\ \hline
  1                         & 0.833 & 0.853 & 0.750 &   -   & 0.000 &   -   & 0.750 & 0.853 & 0.833  \\ \hline
  0                         & 0.888 & 0.960 & 1.000 & 0.888 &   -   & 0.888 & 1.000 & 0.960 & 0.888  \\ \hline
 -1                         & 0.833 & 0.853 & 0.750 &   -   & 0.000 &   -   & 0.750 & 0.853 & 0.833  \\ \hline
 -2                         & 0.666 & 0.533 &   -   & 0.000 & 0.000 & 0.000 &   -   & 0.533 & 0.666  \\ \hline
 -3                         & 0.388 &   -   & 0.000 & 0.000 & 0.000 & 0.000 & 0.000 &   -   & 0.388  \\ \hline
 -4                         &   -   & 0.000 & 0.000 & 0.000 & 0.000 & 0.000 & 0.000 & 0.000 &   -    \\ \hline
\end{tabular}
\end{center}
}
\caption{Numerical estimation of $Z$ in Eq. (\ref{ecopetrol}) for the 
ground state of a Kitaev chain with $\Delta=1$. The data strongly suggests that $Z$ is
rational as long as $w$ and $\mu$ are rational as well. The hyphen 
indicates the parameters for which $Z$ decreases monotonically and slowly
as $N$ grows. Otherwise $Z$ converges for values of $N$ in the range of the tens.}
\label{cervanto}
\end{table}
Interestingly, the numerical values taken by $Z$ in table \ref{cervanto} 
correspond to rational numbers, which allows to fit the data
to the following analytical function
\begin{gather}
Z = \max \left ( \frac{4 |w \Delta| }{(|\Delta| + |w|)^2} \left( 1 - \left( \frac{\mu}{2w} \right)^2 \right),0 \right ).
\label{malta}
\end{gather}
The fact that correlations between the edge sites 
are conditioned by the existence of unpaired MOs is
readily noticeable in this elementary formula.
As can be appreciated, the relation is given in terms of
simple algebraic functions, so that power law coefficients are
rational. The calculation of $Z$ for the first excited
state yields the same ground-state values, while
for the second excited state it seems to give slightly
smaller values.
It remains to be seen whether equation (\ref{malta}) can
be derived entirely by analytical means, as can be expected
due to the integrability of the problem. Analytical results
available so far correspond to the cases {\it i}: $\Delta=w$ 
and {\it ii}: $\mu=0$ \cite{miao1}. Both
instances display structural coincidence with $Z$
in spite of differences with the correlation
measure. This is because terms such as $\hat{c}_1 \hat{c}_N$
and its conjugate do not contribute to the expression 
$\langle i\hat{\gamma}_1 \hat{\gamma}_{2 N} \rangle$ in the 
thermodynamic limit.

The presented evidence hints that end-to-end correlations
are good indicators of the effects generated by edge modes in one
dimension. Due to the topological features of these systems, it is 
reasonable to assume that this result is robust in the presence of 
disorder or interaction. As a consequence, correlations constitute a 
useful inspection mechanism whenever a decomposition in terms of
diagonal modes is not feasible or a topological analysis of the
system's band structure is not practical. Noticeably, the actual
correlation measure seems to be relevant. Entanglement,
which accounts for {\it quantum correlations}, vanishes exponentially 
as $N$ grows due to mixness developed by $\hat{\rho}_{1N}$.
\section{Conclusions}
\label{concl}
Arguments supporting the suitability of end-to-end 
correlations as indicators of unpaired Majorana operators in the
Kitaev chain are discussed. The proposal is verified implementing
a folding protocol in combination with tensor-state representation
to numerically find a given correlation criterion. The results
can be written as a consistent analytical expression that evidence
the connection with Majorana fermions. These findings support
the hypothesis that the same approach can be used in systems with
additional elements like disorder or interaction. Given the characteristics
of the Kitaev chain, it would be interesting to apply similar methodologies
to study Berry phases around the degeneracy points where Majorana fermions
are completely localized.

It is quite likely the folding mechanism employed here have potential applications 
besides the Kitaev chain. First, with some modifications it can be adjusted to 
calculate time evolution or thermodynamic state. Second, it can also be applied 
to chains with long-range hopping or long-range pairing. However, in the way the 
method currently works, it can be used only for integrable models, 
because it is the diagonal modes what is actually folded. It is therefore
desirable to develop a more versatile technique with a broader field of
application. Nonetheless, the protocol can still be useful if interaction
terms are reduced in a mean-field fashion, although it is not known 
how reliable such an approach is. Similarly, it is possible the method has 
applications in the study of open quantum systems and the numerical solution 
of the Lindblad equation.

\section{Acknowledgments}

This research has been funded by Vicerrector\'ia de Investigaciones,
Extensi\'on y Proyecci\'on Social from Universidad del Atl\'antico
under the project 
``Simulaci\'on num\'erica de sistemas cu\'anticos altamente correlacionados''.

\appendix
\section{Tensorial representation of a fermion chain}
\label{tensor}
\begin{center}
\begin{figure}[h]
\begin{center}
\includegraphics[width=0.35\textwidth,angle=-0]{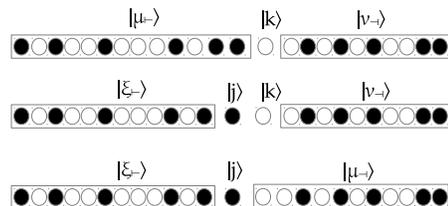}
\end{center}
\caption{The state of the chain can be written in terms of the
Schmidt vectors and a local basis.} 
\label{fig2}
\end{figure}
\end{center}
The reduction scheme presented in the main text can be used to 
write the state as a product of tensors \cite{vidal}.
The basic principle behind such a representation is the use of Schmidt vectors \cite{wiki}  
that emerge in one-dimensional
systems to build basis states that support the global quantum state. Schematically,  
the state of a fermion chain can be represented as in figure \ref{fig2}, using
empty circles for unoccupied sites and black circles for sites with one fermion.
Let us initially focus on one site of the chain, arbitrarily chosen. The Hilbert
space of that site can be expanded using a local basis $|k\rangle$. The elements
of such a basis are $|0\rangle$, to represent an empty site, and $|1\rangle$,
to represent an occupied site. To complement the Hilbert space of the chain,
one can consider the Schmidt vectors covering all the sites to the left of
$|k\rangle$, plus the Schmidt vectors to the right, as shown in the upper
draw of figure \ref{fig2}. As can be seen, such vectors are represented as $|\mu_{\vdash}\rangle$
and $|\nu_{\dashv}\rangle$ respectively. As these vectors are taken as a basis, the
total quantum state is given as a superposition of such vectors, as follows
\begin{gather}
| \psi \rangle = \sum_{\mu} \sum_{\nu}  \sum_{k=0}^1 \lambda_{\mu}  \Gamma_{\mu \nu}^{k} \lambda_{\nu} |\mu_{\vdash} \rangle |k \rangle |\nu_{\dashv} \rangle.
\label{kirko}
\end{gather}
The variables $\lambda_\mu$ and $\lambda_\nu$ are Schmidt coefficients and
as such are real and positive. Although these coefficients can in principle
be absorbed in the definition of the $\Gamma's$, their inclusion is an integral part of the protocol. The 
superposition coefficients are stored in the components of tensor $\Gamma_{\mu \nu}^{k}$.
As a result this tensor is in general complex. Notice that the Schmidt vectors
are orthogonal
\begin{gather}
\langle \mu_{\vdash}  | \mu_{\vdash}' \rangle = \delta_{\mu}^{\mu'} \text{ and } \langle \nu_{\dashv}  | \nu_{\dashv}' \rangle = \delta_{\nu}^{\nu'}.
\end{gather}
If the same expansion is done for each place of the chain, a set of
tensors with no apparent connection among them is created and can serve
as a representation of $|\psi \rangle$. One positive aspect of this
representation is that a local unitary operation like (\ref{twenty-one})
acting on site $l$ has a simple implementation
{\small
\begin{gather}
\hat{U}^{[2l,2l-1]}  | \psi \rangle = \sum_{\mu} \sum_{\nu}
\sum_{k=0}^1 \lambda_{\mu} e^{i \theta_{2l} \left( \frac{1}{2} - k
\right) } \Gamma_{\mu \nu}^{k} \lambda_{\nu} |\mu_{\vdash} \rangle |k \rangle |\nu_{\dashv} \rangle \nonumber \\
 = \sum_{\mu} \sum_{\nu}  \sum_{k=0}^1 \lambda_{\mu} {\Gamma_{\mu \nu}^{k}}' \lambda_{\nu} |\mu_{\vdash} \rangle |k \rangle |\nu_{\dashv} \rangle.
\end{gather}
}
The change involves redefining the tensors as follows
\begin{gather}
{\Gamma_{\mu \nu}^{k}}' = e^{i \theta_{2l} \left( \frac{1}{2} - k \right) } \Gamma_{\mu \nu}^{k}.
\end{gather}
To see how coefficients from different sites relate, let us take vector $|\mu_{\vdash}\rangle$ 
and write it as a product of the local basis, $|j\rangle$, and the Schmidt vectors to the left, $|\xi_{\vdash} \rangle$, 
as indicated in the middle sketch of figure \ref{fig2}. In addition, let us represent this 
expansion  in the following manner
\begin{gather}
| \mu_{\vdash} \rangle = \sum_{\xi} \sum_{j=0}^1 \lambda_\xi \Gamma_{\xi \mu}^j  |\xi_{\vdash} \rangle |j \rangle.
\end{gather}
Replacing this expression in (\ref{kirko}) it results 
{\footnotesize
\begin{gather}
| \psi \rangle = \sum_{\xi} \sum_{\nu} \sum_{j=0}^1 \sum_{k=0}^1
\lambda_\xi \left ( \sum_{\mu}  \Gamma_{\xi \mu}^j  \lambda_{\mu}
\Gamma_{\mu \nu}^{k} \right ) \lambda_{\nu}  |\xi_{\vdash} \rangle |j \rangle  |k \rangle |\nu_{\dashv} \rangle  =  \nonumber \\  
 \sum_{\mu} \lambda_{\mu} \left( \sum_{\xi} \sum_{j=0}^1  \lambda_\xi  \Gamma_{\xi \mu}^j |\xi_{\vdash} \rangle |j \rangle \right) \left( \sum_{\nu}  \sum_{k=0}^1 \Gamma_{\mu \nu}^{k}  \lambda_{\nu}  |k \rangle |\nu_{\dashv} \rangle  \right).
\label{twenty-six}
\end{gather}
}
In the last expression the chain has been divided as a
Schmidt decomposition with Schmidt coefficients $\lambda_\mu$. This
implies that the set of vectors
\begin{gather}
| \mu_{\dashv} \rangle = \sum_{\nu} \sum_{k=0}^1  \Gamma_{\mu \nu}^k \lambda_\nu |k \rangle |\nu_{\dashv} \rangle,
\end{gather}
must be a set of Schmidt vectors to the right, making $\langle \mu_{\dashv}  | \mu_{\dashv}' \rangle = \delta_{\mu}^{\mu'}$.
The Schmidt decomposition of the state can then be written in the familiar form
\begin{gather}
| \psi \rangle = \sum_{\mu}  \lambda_\mu | \mu_{\vdash} \rangle  |\mu_{\dashv} \rangle. 
\end{gather}
Let us now consider an unitary operation acting on consecutive sites $l$ and $l+1$, as 
for instance transformation (\ref{twenty-two}). The operation can be represented as
{\footnotesize
\begin{gather}
\hat {U}^{[2l+1,2l]} | \psi \rangle = \sum_{\xi} \sum_{\nu} \sum_{J=0}^1
\sum_{K=0}^1 \nonumber \\
\left ( \lambda_\xi \lambda_{\nu} \sum_{j=0}^1 \sum_{k=0}^1  U_{JK,jk}   \sum_{\mu}  \Gamma_{\xi \mu}^j  \lambda_{\mu}  \Gamma_{\mu \nu}^{k} \right )   |\xi_{\vdash} \rangle |J \rangle  |K \rangle |\nu_{\dashv} \rangle.      
\label{twenty-five}
\end{gather}
}
The resulting expression is no longer an evident Schmidt decomposition  
but it can be rearranged as one in the next manner. Let us write the operation 
in parenthesis as
\begin{gather}
\lambda_\xi \lambda_\nu \sum_{j=0}^1 \sum_{k=0}^1 \sum_{\mu} U_{JK,jk}
\Gamma_{\xi \mu}^j  \lambda_{\mu} \Gamma_{\mu \nu}^{k} = M_{\xi J, K
\nu} = M_{\alpha,\beta} \nonumber
\end{gather}
In the last step the pairs of indices $(\xi, J)$ and $(K, \nu)$ have been replaced
by single indices $\alpha$ and $\beta$. Notice that grouping indices is essentially
a notation change. It resorts to the possibility of joining Hilbert spaces
from adjacent sections of the chain. Matrix $M_{\alpha,\beta}$ has no restrictions apart
from normalization. It is in general complex  and is not necessarily square.
Such a matrix can be written as a product of (less arbitrary) matrices applying a
singular value decomposition (SVD) \cite{wiki}
\begin{gather}
M_{\alpha,\beta} = \sum_{\alpha'} \sum_{\beta'} T_{\alpha,\alpha'} \Lambda_{\alpha',\beta'} T_{\beta',\beta}.
\label{twenty-four}
\end{gather}
Both $T_{\alpha,\alpha'}$ and $T_{\beta',\beta}$ (different matrices) are complex and unitary,
their rows (or columns) being orthogonal vectors. They are also square matrices.
Matrix $\Lambda_{\alpha',\beta'}$ is real and diagonal.
\begin{gather}
\Lambda_{\alpha',\beta'} =
\left(
\begin{array}{cccc}
\lambda_1 & 0 & 0 & ... \\
0 & \lambda_2 & 0 & ... \\
0 & 0 & \lambda_3 & ... \\
\vdots & \vdots & \vdots & \ddots
\end{array}
\right)
\end{gather}
Normalization requires $\lambda_1^2 + \lambda_2^2 + \lambda_3^2 +... = 1$. All the
$\lambda$'s are positive. In many numerical applications the number of $\lambda$'s
is artificially fixed. Here the number of coefficients is handled dynamically
and only those below numerical precision are discarded. 

The double sum in (\ref{twenty-four}) can be reduced to a single sum
\begin{gather}
M_{\alpha,\beta} = \sum_{\mu'} T_{\alpha,\mu'} \lambda_{\mu'} T_{\mu',\beta}.
\end{gather}
One can in addition write the labels $\alpha$ and $\beta$ in terms of the 
original labels
\begin{gather}
M_{\alpha,\beta} = \sum_{\mu'} T_{\xi J,\mu'} \lambda_{\mu'} T_{\mu',K \nu} = \sum_{\mu'} \lambda_\xi \Gamma_{\xi \mu'}^J \lambda_{\mu'} \Gamma_{\mu' \nu}^K \lambda_\nu.
\end{gather}
In the last step the components of the $T$'s have been renamed. Notice that
the $\Gamma$'s in the last sum are different from the ones appearing in the initial state.
No emphasized distinction is made in order not to overload the notation, but 
tensors with $\mu'$ are all new. Also notice that neither $\lambda_\xi$ nor $\lambda_\nu$ have
changed. As $J$ and $K$ are integer labels without explicit meaning, it is valid to rename
them with their lower-case equivalents $j$ and $k$. Introducing the final expression
in (\ref{twenty-five}) gives 
{\scriptsize
\begin{gather}
\hat {U}^{[2l+1,2l]} | \psi \rangle = \sum_{\xi} \sum_{\nu} \sum_{j=0}^1
\sum_{k=0}^1   \lambda_\xi \left (  \sum_{\mu}  \Gamma_{\xi \mu'}^j
\lambda_{\mu'}  \Gamma_{\mu' \nu}^{k} \right ) \lambda_{\nu}
|\xi_{\vdash} \rangle |j \rangle  |k \rangle |\nu_{\dashv} \rangle.
\nonumber    
\end{gather}
}
The state is in ``canonical form'', i.e., written with respect to the (new) Schmidt
vectors of the chain, since the states formed as
\begin{gather}
| \mu_{\vdash}' \rangle = \sum_{\xi} \sum_{j=0}^1 \lambda_\xi \Gamma_{\xi \mu'}^j  |\xi_{\vdash} \rangle |j \rangle,
\end{gather}
and 
\begin{gather}
| \mu_{\dashv}' \rangle = \sum_{\nu} \sum_{k=0}^1  \Gamma_{\mu' \nu}^k \lambda_\nu |k \rangle |\nu_{\dashv} \rangle,
\end{gather}
are orthogonal and normalized because they are the entries of matrices $T_{\alpha,\alpha'}$
and $T_{\beta',\beta}$ respectively. 

This representation is very convinient to calculate
local mean values. Using (\ref{kirko}) it can be shown that the reduced density matrix of 
a given site is
\begin{gather}
\hat{\rho}_{k , k'} = \sum_{k=0}^1 \sum_{k'=0}^1 \sum_\mu \sum_\nu \lambda_\mu^2 \lambda_\nu^2 \Gamma_{\mu \nu}^k  \Gamma_{\mu \nu}^{k' *}| k \rangle \langle k' |.
\end{gather}
A mean value corresponding to a matrix $\hat{\tau}$ that operates exclusively on that
site can be found as 
\begin{gather}
\langle \hat{\tau} \rangle = Tr(\hat{\rho} \hat{\tau}).
\end{gather}
One can work in an analogous way in the space of two consecutive positions using
the corresponding reduced density matrix. It can be shown that this matrix
can be written as
\begin{gather}
\hat{\rho}_{ j k, j' k' } = \sum_{j k} \sum_{j' k'} \sum_{\xi} \sum_{\nu}   {}_{\xi}^{j}Y_{\nu}^{k} \text{ } {}_{\xi}^{j'}Y_{\nu}^{ k'\text{} *}  |j k\rangle \langle j' k'|,
\end{gather}
such that 
\begin{gather}
{}_{\xi}^{j}  Y_{\nu}^{k} = \lambda_{\xi} \lambda_{\nu}  \sum_{\mu}  \Gamma_{\xi \mu}^{j} \lambda_\mu \Gamma_{\mu \nu}^k.
\end{gather}
Sometimes it is also useful
to know how to obtain the state coefficients in the Fock basis in
terms of this tensor representation. Such a relation can be derived following the
arguments in reference \cite{vidal}, thus yielding
\begin{gather}
c_{k_1 k_2 ... k_N} = \sum_{\mu} \sum_{\nu} ... \sum_{\xi} \Gamma_{1 \mu }^{k_1} \lambda_\mu \Gamma_{\mu \nu }^{k_2} \lambda_\nu ... \lambda_\xi \Gamma_{\xi 1 }^{k_N}.
\end{gather}
These operations can be efficiently computed if the number of Schmidt
coefficients involved is not too large.
\subsection{Example}
Let us initially consider a chain with no fermions. In the Fock basis the
state is given by
\begin{gather}
|\psi \rangle = |000...0\rangle.
\end{gather}
When this state is split in adjacent parts, the corresponding Schmidt decomposition is trivial:
There is one vector to the left, one vector to the right and the only Schmidt coefficient
is $1$. From this observation the canonical decomposition can be built directly
\begin{gather}
\lambda = 1, \text{ } \Gamma_{1 1}^0 = 1, \text{ } \Gamma_{1 1}^1 = 0.
\end{gather}
The same pattern repeats for every place of the chain. Now let us consider the following
unitary operation 
\begin{gather}
\hat{U} = 
\frac{1}{\sqrt{2}}
\left (
\begin{array}{cccc}
 1 & 0 & 0 & i \\
 0 & 1 & i & 0 \\
 0 & i & 1 & 0 \\
 i & 0 & 0 & 1  
\end{array}
\right ).
\end{gather}
For simplicity let us suppose that $\hat{U}$ acts on the first two places.
The action of this operator on the state is
\begin{gather}
\hat{U}^{[3]} |\psi \rangle = \frac{1}{\sqrt{2}} \left( |00 \rangle + i|11\rangle \right) |0...0\rangle.
\end{gather}
To build a canonical decomposition (the canonical decomposition is not unique),
one sees the state as a composition of a local basis plus the Schmidt vectors to
the right and left. Taking the local basis of the first site, the state can be
written as
\begin{gather}
\hat{U}^{[3]} |\psi \rangle = \frac{1}{\sqrt{2}} \left( |0 \rangle |00...0\rangle  + i|1 \rangle |10...0\rangle \right).
\end{gather}
Vectors $|\nu_1\rangle = |00...0\rangle$ and $|\nu_2\rangle = |10...0\rangle$ are
normalized and orthogonal, therefore, they are valid Schmidt vectors. 
In this form it is possible to read out the canonical coefficients, finding
\begin{gather}
\lambda_1^{[1]} = \frac{1}{\sqrt{2}}, \text{ }  \lambda_2^{[1]} = \frac{1}{\sqrt{2}}, \\
\Gamma_{1 1}^{0[1]} = 1, \text{ } \Gamma_{1 1}^{1[1]} = 0,  \\
\Gamma_{1 2}^{0[1]} = 0, \text{ } \Gamma_{1 2}^{1[1]} = i.
\end{gather}
The superscript $[1]$ is added to emphasize that these coefficients
correspond to the first site. With respect to this decomposition, the 
estate can be visualized in the following manner (with the superscript omitted) 
\begin{gather}
\hat{U}^{[3]} |\psi \rangle =  \lambda_1 \Gamma_{1 1}^0 |0 \rangle |\nu_1 \rangle  + \lambda_2 \Gamma_{1 2}^1 |1 \rangle |\nu_2\rangle.
\end{gather}
This case has been sufficiently simple to allow a direct
determination of the canonical representation. In other circumstances
the protocol presented in the previous section can be used to
build a representation in accordance with the original proposal using
a systematic approach.
\section{Reduced density matrix of the chain ends}
\label{rdm}
To find the reduced density matrix the state is written as a tensor 
product making explicit reference to the components of each site
\begin{gather}
|\psi \rangle =  \sum_{\alpha_1,\alpha_2,...,\alpha_{N-1}} | \alpha_0 \alpha_1 \rangle^{[1]}  | \alpha_1 \alpha_2 \rangle^{[2]}... \nonumber \\
|\alpha_{N-2} \alpha_{N-1}\rangle^{[N-1]} |\alpha_{N-1} \alpha_{N}\rangle^{[N]},
\end{gather}
where 
\begin{gather}
|\alpha_k \alpha_l \rangle^{[n]} =  \sum_{j} \Gamma_{\alpha_k \alpha_l}^{j[n]} \lambda_{\alpha_l}^{[n]} |j\rangle.
\label{mist}
\end{gather}
The superscript in square brackets is used to specify the position
in the chain associated to the corresponding tensor. Such a superscript
is dropped in the subsequent development to simplify the notation.
The reduction can be effectuated by bracketing corresponding spaces
{\scriptsize
\begin{gather}
\hat{\rho}_{1N} = Tr_{\{2,3...,N-1\}}( |\psi \rangle  \langle \psi | ) = \nonumber \\
 \sum_{ \binom{\alpha_1,\alpha_2,...,\alpha_{N-1}}{{\alpha'}_1,{\alpha'}_2,...,{\alpha'}_{N-1}}
}   |\alpha_0 \alpha_1 \rangle \langle {\alpha_0' \alpha'}_1 | \langle \alpha_1' \alpha_2' |{\alpha}_1 {\alpha}_2 \rangle   
\langle \alpha_2' \alpha_3' |{\alpha}_2 {\alpha}_3 \rangle \nonumber  \\
 ...\langle \alpha_{N-2}' \alpha_{N-1}' |{\alpha}_{N-2} {\alpha}_{N-1} \rangle  |\alpha_{N-1} \alpha_{N} \rangle \langle {\alpha'}_{N-1}  \alpha_{N}' |. \nonumber 
\end{gather}
}
The above expression can also be written as a concatenation of index contractions
{\scriptsize
\begin{gather}
\sum_{\alpha_1,{\alpha'}_1,\alpha_{N-1},{\alpha'}_{N-1}} |\alpha_0 \alpha_1 \rangle \langle {\alpha_0' \alpha'}_1 |  M_{ \{ \alpha_1 {\alpha'}_1 \}  \{ \alpha_2 {\alpha'}_2 \}}  M_{ \{ \alpha_2 {\alpha'}_2 \}  \{ \alpha_3 {\alpha'}_3 \} } \nonumber \\
... M_{ \{ \alpha_{N-2} {\alpha'}_{N-2} \}  \{ \alpha_{N-1} {\alpha'}_{N-1} \} } |\alpha_{N-1} \alpha_{N} \rangle \langle {\alpha'}_{N-1}  \alpha_{N}' |. \nonumber 
\end{gather}
}
Thus, it all can be written as a single connecting matrix
{\scriptsize
\begin{gather}
\hat{\rho}_{1N} = \sum_{\alpha_1,{\alpha'}_1,\alpha_{N-1},{\alpha'}_{N-1}}   |\alpha_0 \alpha_1 \rangle \langle { \alpha_0' \alpha'}_1 | M_{ \{ \alpha_1 {\alpha'}_1 \}  \{ \alpha_{N-1} {\alpha'}_{N-1} \} } \nonumber \\
...|\alpha_{N-1}  {\alpha}_{N} \rangle \langle {\alpha'}_{N-1} {\alpha'}_{N} |. \nonumber
\end{gather}
}
The calculation of $M_{ \{ \alpha_1 {\alpha'}_1 \}  \{ \alpha_{N-1} {\alpha'}_{N-1} \} }$ 
unavoidably involves all the tensors in the bulk of the representation
and is the numerically heaviest task of the procedure.
The resulting expression is a $4\times4$ matrix in the Fock basis
of the chain edges.
\end{document}